\documentclass[doublespacing]{elsart}

\usepackage{amssymb}
\usepackage{graphicx}
\begin{document}

\begin{frontmatter}



\title{SuperAGILE: the hard X-ray Imager for the AGILE space mission}


\author{M. Feroci$^{*}$, E. Costa, P. Soffitta, E. Del Monte, G. Di Persio,}
\author{I. Donnarumma, Y. Evangelista, M. Frutti, I. Lapshov$^{1}$,}
\author{F. Lazzarotto, M. Mastropietro$^{2}$, E. Morelli$^{3}$, L. Pacciani, }
\author{G. Porrovecchio, M. Rapisarda$^{4}$, A. Rubini, M. Tavani, and A. Argan }

\address{INAF - Istituto di Astrofisica Spaziale e Fisica Cosmica - Roma}
\address{$^{1}$also, IKI, Academy of Sciences, Moscow, Russia}
\address{$^{2}$also, CNR - Istituto Metodologie Inorganiche e dei Plasmi - Area
Ricerca Montelibretti}
\address{$^{3}$INAF - Istituto di Astrofisica Spaziale e Fisica Cosmica - Bologna}
\address{$^{4}$also, ENEA - Centro Ricerche - Frascati}
\address{$^{*}$ Corresponding Author: INAF/IASF - Via Fosso del Cavaliere 100
00133 Rome, Italy - e-mail: marco.feroci@iasf-roma.inaf.it - ph.
+39-06-4993-4099, fax +39-06-2066-0188}

\begin{abstract}
SuperAGILE is a coded mask experiment based on silicon microstrip
detectors. It operates in the 15-45 keV nominal energy range,
providing crossed one-dimensional images of the X-ray sky with an
on-axis angular resolution of 6 arcmin, over a field of view in
excess of 1 steradian. It was designed as the hard X-ray monitor
of the AGILE space mission, a small satellite of the Italian Space
Agency devoted to image the gamma-ray sky in the 30 MeV - 50 GeV
energy band. The AGILE mission was launched in a low-earth orbit
on 23$^{rd}$ April 2007. In this paper we describe the SuperAGILE
experiment, its construction and test processes, and its
performance before flight, based on the on-ground test and
calibrations.
\end{abstract}

\end{frontmatter}

\section{Introduction}
\label{intro}

Gamma-ray astronomy is one of the least explored windows on our
Universe. The intrinsic instrumental difficulty to detect and
image photons with energy between few tens of MeV and tens of GeV
has led to only two narrow observational periods so far, with
SAS-2 (1973, Fichtel et al. 1975) and Cos-B (1975-1982, Bennet
1990), and with EGRET (1991-1999, Hartman et al. 1999). Originally
proposed and designed to fill the gap between EGRET and the
larger, GLAST mission\footnote{e.g.,
http://www-glast.stanford.edu/}, expected to become operational in
2008, due to significant delays the Italian AGILE gamma-ray
mission is expected to observe the gamma-ray sky over the same
period of time, but with a different observing strategy (Tavani et
al. 2006). In addition to the observation of the "known" gamma-ray
sources, the AGILE main scientific goal is the discovery of new
gamma ray transients (e.g., gamma ray bursts, outbursts of AGNs,
microquasars, ...) and of the counterparts of the unidentified
EGRET sources (e.g., Grenier 2003). To accomplish this, the
gamma-ray imager on board AGILE has a very large field of view
(approximately 2 steradians) and a very good effective area and
angular resolution at large off axis angles.

AGILE (Tavani et al. 2006), the first small mission of the Italian
Space Agency, was launched from the Shriharikota base in India
with a PSLV launcher\footnote{e.g., http://www.isro.org/pslv.htm}
on 23$^{rd}$ April 2007 in an equatorial orbit
($\sim$2$^{\circ}$.5 inclination) at $\sim$545 km altitude. The
$\sim$350 kg AGILE satellite carries a $\sim$120 kg scientific
payload composed of a Gamma Ray Imaging Detector (GRID) - a
silicon microstrip tracker and a CsI "mini" calorimeter,
surrounded by a plastic anticoincidence - and a hard X-ray
monitor, SuperAGILE. In fact, despite the improved efficiency of
the gamma-ray imager with respect to its predecessors, the
observation of celestial gamma-ray sources suffers severe photon
starvation and angular confusion. The simultaneous observation of
the gamma-ray sources with an instrument operating in hard X-rays,
an energy range where higher sensitivity and better angular
resolution can be achieved, helps with the identification of the
source, with the very important benefit of allowing to correlate
and study the simultaneous emission in two distant energy bands.
This is indeed the basic scientific justification for having an
experiment like SuperAGILE onboard the gamma-ray mission AGILE.

SuperAGILE is a coded mask experiment, operating in the nominal
energy range 15-45 keV, imaging a field of view in excess of 1
steradian with a 6 arcminutes one-dimensional angular resolution
(on axis), in two orthogonal directions in the sky. The main
scientific goal of SuperAGILE is to provide a prompt, unambiguous
and accurate identification of the gamma-ray sources observed with
the GRID. Based on what we know today, this is a task that can be
accomplished on a relatively small sample of sources (e.g.,
gamma-ray pulsars, blazars, gamma-ray bursts). However, with a
sensitivity of $\sim$15 mCrab (in 50 ks of observation) and field
of view of 107$^{\circ} \times$68$^{\circ}$ (at zero response),
SuperAGILE will also operate as a (quasi) all sky monitor of the
X-ray sky, allowing the monitoring of known Galactic sources (low
and high mass X-ray binaries, magnetars, ...) and possibly the
discovery of new transient sources.

This paper describes the concept and details of the design and
construction of the flight model of the SuperAGILE experiment,
providing the reference performance and calibrations obtained
before launch. A Detailed description of the ground calibration
procedures and results will be given in a future publication.

\section{The Concept and the Context}
\label{concept}

SuperAGILE (SA) was conceived as a monitoring instrument for the
primary mission experiment, GRID. As such, the first requirement
was to use minimal resources and do not significantly affect the
scientific response of GRID. This was a major constraint to the SA
design and performance. The operating principle of GRID is based
on the conversion of gamma-ray photons in electron-positron pairs
by thin tungsten sheets, and tracking the pairs, in energy and
direction, by microstrip silicon detectors, thus reconstructing
the kinematics of the impinging photon.

Since to absorb X-rays a thin amount of matter is sufficient, the
only possible location for SA, compatible with the small dimension
of the AGILE payload and with the scientific request of a
co-aligned field of view (FOV) for the two instruments, is on top
of GRID. Thus, any mass of the SA experiment will itself acts as
an unwanted converter of gamma-ray photons, resulting in the
decrease of the GRID efficiency and in the increase of its
particle background. This brought to the use of materials with low
atomic number and keep the weight at minimum, to decrease the
conversion probability for photons. The residual (average)
probability of SA to convert a gamma-ray photon, thus decreasing
the signal to noise of GRID, was then kept at less than 10\%. For
the same reasons, SA had to be located inside the anticoincidence
surrounding the GRID. This implies that a 5 mm thick layer of
plastic scintillator, with its support structure, completely
obstructs the SA FOV, largely decreasing its efficiency,
especially at energies lower than 20 keV (e.g., its transparency
on axis is 70\% at 15 keV).

SA had then to satisfy the scientific requirements relevant to its
monitoring function. First, covering the largest fraction of the
GRID FOV ($\sim$2 steradian) was obtained by co-aligned FOVs.
Then, a trade-off was necessary on the size of the FOV: contrary
to the gamma-ray band, at X-rays the main source of background for
an instrument with a large FOV are X-ray photons of the isotropic
diffuse X-ray background, undistinguishable from the good source
photons, at detection level. Thus, the larger is the FOV, the
lower is the experiment sensitivity and the larger is the
telemetry required to transmit individual events to ground. The
target sensitivity was therefore set to $\sim$10-15 mCrab 5
$\sigma$ in one day (1 mCrab corresponds to
$\sim$4$\times$10$^{-4}$ photons cm$^{-2}$ s$^{-1}$ or
$\sim$1.5$\times$10$^{-11}$ erg cm$^{-2}$ s$^{-1}$ in 15-45 keV),
a level adequate to the arcmin-level angular resolution, with
respect to the source confusion limit. Taking into account the
fixed geometry of the detectors, the choice for the FOV is a
rectangle, 107$^{\circ} \times$68$^{\circ}$ at zero response, for
each pair of detectors. The two pairs of detectors are then
oriented at approximately 90$^{\circ}$, so that SA can provide the
two coordinates of sources, when occurring within the central
68$^{\circ}$x68$^{\circ}$ part of the FOV, and still provide the
X-ray detection of a gamma-ray source and its single coordinate
when at the edges of the FOV. The mask-detector distance was
instead chosen such that the source location accuracy is in the
range of few arcminutes, thus useful to follow-up the SA
localizations with other observatories, at X-rays and other
wavelengths. On the other hand, the systematic error in the
satellite attitude reconstruction is expected in the range 1-2
arcminutes, thus we expect to reach the systematics for most
sources.

Finally, to keep the development, manufacturing and qualification
costs at minimum, SA employs the same constructive elements as the
GRID. In particular, the silicon detectors, the ASIC model, the
high voltage power supply, the carbon fiber detector tray
technology are just the same as the GRID.

Figure \ref{sa_structure} shows a sketch with an exploded view of
the SA experiment, separating the detection plane, the collimator
and the mask. In Table \ref{sa_char} the main instrument
parameters are summarized. The development of the SA flight model
was preceded by a series of tests on prototypes and by the
construction and test of an engineering model.  Details on them
may be found in Soffitta et al. (2002) and Feroci et al. (2004)
and references therein.

\begin{table}[h!]
\centering
 \caption{Main instrumental parameters of the SuperAGILE experiment. }
\bigskip
\label{sa_char}

\begin{tabular}{|c|c|}
  \hline
  Parameter                 & Value   \\
                            &  \\
  \hline
  Detector                  & n-bulk, p-strip, single-sided, AC-coupled Si
  $\mu$strip \\
  Detector thickness        & 410 $\mu$m \\
  Nominal Energy Range      &  15-45 keV  \\
  Energy Resolution         & $\sim$8 keV FWHM  \\
  Geometric Area            & 1344 cm$^{2}$  \\
  Effective Area (max)      &  230 cm$^{2}$  \\
  Field of View             & 107$^{\circ} \times$68$^{\circ}$   \\
  Mask Transparency         & 50\% \\
  Angular Resolution        & 6 arcmin   \\
  Mask Element Size         & 242 $\mu$m  \\
  Detector Strip Size       & 121 $\mu$m $\times$ 190 mm   \\
  Mask-Detector Distance    & 142 mm   \\
  Timing Resolution         & 2 $\mu$s   \\
  Timing Accuracy           & 5 $\mu$s  \\
  Point Source Location Accuracy & 1-2 arcmin    \\
  Point Source Sensitivity  & 15 mCrab (1D)    \\
  Data Transmission         & Event-by-Event, 32-bit/event   \\
  Telemetry Allocation      & 25 kbit/s  \\
 \hline
\end{tabular}
\end{table}

\section{The Detection Plane}
\label{dp}

The Detection Plane (DP) is the sensitive unit of SA, composed of
the detector tray on which the SA Detection Units (DUs, silicon
detectors plus front-end electronics) are fixed.

The detector tray is a 44 cm x 44 cm x 0.5 cm support plane made
of aluminum honeycomb, covered with two outer layers of carbon
fiber tissue. To prevent X-ray background photons to reach the
detectors from below, in the area where the GRID material does not
ensure enough shielding, a 120 $\mu$m thick tungsten frame was
glued to the back side of the tray. The total weight of the tray
is $\sim$850 g. A special care was devoted to its planarity,
required by the gluing process of the DUs, that was requested and
obtained within 0.15 mm overall. The detector tray was
manufactured by Plyform (Italy) and Oerlikon Contraves Italia.

\subsection{The Silicon Microstrip Detectors}
\label{detectors}

 The sensitive elements of the SA experiment are
single-sided, AC-coupled, n-type, p-strip silicon microstrip
detectors manufactured by Hamamatsu (Japan). Each of the 4 SA
detectors is composed of 4 Silicon tiles, 95 mm x 95 mm x 0.41 mm
in size, implanted with a pattern of 768 60$\mu$m wide metal
strips at 121 $\mu$m pitch. The strip pattern is surrounded by a
guard ring and a bias ring. The active area of each tile is thus
approximately 84 cm$^{2}$, providing an active area of about 336
cm$^{2}$ per detector. Each detector was assembled first by gluing
two tiles head-to-head to form one ladder. Then, two ladders were
independently glued side by side, with a controlled separation,
onto a Cu/Au grid-coated kapton foil (hereafter "flexbias"), also
used to bring the electrical positive bias ($\sim$90 V) from the
SAFEE (SuperAGILE Front End Electronics) boards to the Si
detectors. The electrical contact between the back of the Si tiles
and the flexbias is guaranteed by 5 spots of a conductive epoxy
glue per tile, while the mechanical stiffness of the system is
provided by a larger number of spots of silicon glue.

The mechanical assembly was performed giving special care to the
alignment issues. The head-to-head assembly was such that the
individual strips are phased within 5$\mu$m and parallel to better
than 10 arcsec. The mechanical distance between the two ladders
was then fixed such that the separation between the innermost
active strips of the two ladders is equivalent to the smallest
integer number (18) of strip pitches, in order to keep the impact
of the mechanical separation on the coded imaging capability at
minimum. This goal was achieved to within $\sim$10 $\mu$m, also
thanks to the excellent mechanical quality of the mask and cuts of
the Hamamatsu silicon tiles.

 The strips of the two tiles composing one ladder are bonded
one-to-one by 25 $\mu$m Aluminum wires, so that effective 19-cm
long strips are obtained, whose capacitive load is 30 pF. Each of
these strips is connected to the bias ring through a parallel of
two 40 M$\Omega$ on-chip poly-resistors. Bias and guard rings are
grounded.

Further details about the detector assembly procedure, carried out
by the MIPOT company (Italy), may be found in Rapisarda et al.
(2004).

\subsection{The SuperAGILE Front End Electronics (SAFEE)}
\label{safee}

Each of the 4 SA detectors is read-out through an independent
SuperAGILE Front-End Electronics (SAFEE) unit. The "core" of the
SAFEE are its 12 XAA1.2 ASIC chips, each of them including 128
independent spectroscopic chains, composed of preamplifier,
discriminator, amplifier and peak stretcher. Each detector strip
is connected to one of the 128 channels of one XAA1.2. The pitch
of the detector (121 $\mu$m) was adapted to the 96 $\mu$m pitch of
the XAA1.2 by means of a pitch adapter (gold tracks deposited on a
two layer alumina board). Connections between detector strips and
pitch adapter and between the control pads of the ASIC and the
SAFEE PCB are made by 25 $\mu$m Al wire bondings. Connections
between the pitch adapter and the ASIC input channels are made by
17 $\mu$m Al wires.

A detailed description of the SAFEE is given in Pacciani et al.
(2007), whereas the XAA1.2 characteristics and performances may be
found in Del Monte et al. (2007). Here we just summarize the major
features of these systems.

Each SAFEE is both conceptually and electrically organized in 4
almost independent Daisy Chains (DCs), including 3 ASIC dies each.
When in DC, the chain of three XAA1.2 chips acts as it was a
single chip. In fact the 12 biases requested to configure each
XAA1.2 chip are in common for the 3 elements in one DC, with a
"winner takes it all" logic (meaning that a triggered channel
disables all the others until after complete processing) applies
to the 384 channels of a DC. Upon trigger, the relevant DC places
on its output bus the following analog information, in
differential current signals: multiplicity signal (allows to
select triggers from single and multiple strips), amplitude, group
address (most significant bits of the detector strip address,
between 1 and 12) and strip address (least significant bits,
between 1 and 32). These signals are directly passed to the
144-pin connector interfacing the SAFEE to its SAIE (SuperAGILE
Interface Electronics, see \S\ref{saie}). The dead time of each DC
depends on the setting of three parameters of the XAA1.2,
determining the delay from the discrimination to the start of the
"output-ready" gate (TrigDelBias), the duration of the gate
(TrigWidthBias) and the time to the chip reset (ResWidthBias). At
the AGILE launch the total deadtime per DC was set as about 80
$\mu$s, with few ($\sim$5-10) microseconds variations from channel
to channel and $\sim$10$\mu$s from ASIC to ASIC. This long
"analogue" deadtime makes it negligible the processing time of the
A/D converter in the SAIE, that runs in parallel to the chip reset
time. It is worth noticing that the event processing from the 16
SA DCs is independent and multiplexed in the SAIEs, thus the
overall instrument operational deadtime can be conventionally
defined as 1/16 of that of the individual DC, that is $\sim$5
$\mu$s.

Only 8 of the 12 configuration biases of the XAA1.2 were made
programmable by telecommand. To this purpose, 4 DAC octal chips
were used, one per DC, whose reference voltage is provided by the
regulated $\pm$2 V lines powering the ASICs themselves. In this
set-up, the biases are automatically dumped off at the power
supply switch off.

The dimension of the SA tray were constrained by external
parameters (e.g, GRID design and internal dimensions of the
anticoincidence subsystem). Therefore the electronic components
that were required for a SAFEE could not all fit on a single
printed circuit board (PCB) lying on the same plane as the
detector. Still, the ASICs had to be wire-bonded to the detector
strips and lay close to them. Therefore, the SAFEE was designed as
a 6-layer rigid-flex PCB, with two rigid parts connected by a
flexible part. The part hosting the ASICs and the bias resistors
was placed horizontally near to the detector, while the DACs and
their circuits are hosted on a larger rigid part, screwed to the
external wall of the collimator (see photo in Figure \ref{du}).
The vertical SAFEE is protected by a custom aluminum box.

Due to the required density of the SAFEE circuit and to the
low-cost approach of the AGILE mission, most of the electronic
components - including the XAA1.2 - are commercial grade, except
for resistors that are MIL-standard. In order to ensure a minimum
level of quality assurance, a few environmental tests were carried
out. In particular, the XAA1.2 were radiation tested for Latch-up,
Single Event Upset (SEU) and Dose effects at the ion accelerator
facility of INFN in Legnaro (Del Monte et al., 2005). The DACs
(Analog Device's AD8842AR) were tested for latch-up and SEU by the
Hirex Engineering company (France). Finally, each complete SAFEE
underwent a 10-day burn-in test at 75$^{\circ}$C. Further
environmental tests (thermal-vacuum, acoustic, EMC, vibrations)
were carried out at higher levels of system integration (detection
plane, experiment, payload and satellite). All of the above tests
were successful, and from them no warning was derived for the
operations in orbit. However, for further safety, the power
supplies lines of both the XAA1.2 ($\pm$2V) and the AD8842AR
($\pm$5V) were equipped with autonomous latch-up protection
circuits onboard.

The power consumption of the XAA1.2 was one of the issues of the
ASIC optimization (the XA1.3 version used $\sim$3 mW/channel). In
our baseline configuration, the XAA1.2 requires $\sim$1
mW/channel. Changes to the setting of the biases controlling the
analog signal shape can bring to changes to this value. A back-up
solution is always represented by the exclusion of the stretcher
circuit, allowing a decrease in power consumption by a factor more
than 2. Considering the consumption on the 5 V and 130 V lines,
the total power requested by the entire SA DP (excluding only the
SAIE boards described below) is less than 7 W overall. Further
details may be found in Pacciani et al. (2007) and Del Monte et
al. (2007).

\subsection{Integration and Alignment of Detection Units}
\label{du_int}

One SA Detection Unit (DU) is composed of one detector (that is, 4
Silicon tiles), one SAFEE and their connecting Cu/Au coated kapton
board, the so-called "flexbias". The horizontal section of the
SAFEE board is glued on the flexbias already connected to the
detector by conductive glue. The shape of the two is such that the
high voltage contact on the SAFEE board can be brought, by
soldering, to the flexbias and to the two individual ladders of
the detector thereafter. The mechanical connection between the
detector and the SAFEE PCB is then reinforced with head-to-head
spots epoxy gluing. Finally, the detector strips and SAFEE input
channels are interconnected through 25$\mu$m/17$\mu$m aluminum
wire bonding, thus completing a SA Detection Unit.

\subsection{Integration and Alignment of the Detection Plane}
\label{dp_int}

After proper testings (see \S\ref{perf}), the 4 SA DUs were
aligned and integrated on the detector tray. For the alignment of
the experiment the following approach was used: have the internal
alignment of the masks and DUs systems to better than half of the
overall alignment requirement and leave only the degree of freedom
of aligning the two complete systems one to the other. The
scientific requirement on the overall detector-mask alignment was
set as half of detector strip (60 $\mu$m) over its total length
(190 mm) (corresponding to an angle of $\sim$60 arcsec). In fact,
the main effect of a mask-detector misalignment is the loss in the
signal-to-noise ratio of the reconstructed source (e.g.,
Charalambous et al. 1984), due to the fraction of the encoded
photons detected by the "wrong" detector strip. In practice, this
effect becomes important (that is, of the order of few percent)
when the misalignment is larger than half of the mask element
width. For SA, this condition implies a linear displacement of 121
$\mu$m, thus our requirement of 60 $\mu$m corresponds to half of
the error budget on the alignment.

The DUs were spot-glued directly onto the carbon fiber detector
tray. As a consequence, their alignment had to be obtained
simultaneously to their integration. An accurate mechanical
reference system was obtained by tightly inserting 3 steel pins
with very precise cross-shaped engravings (allowing a resolution
of $\sim$5 $\mu$m) in the outer frame of the tray. From this point
onward, these 3 points represented the "absolute" reference frame
of the SA DP, in which the ideal geometrical positions of each DU
were predetermined. Each DU, supported by a vacuum plier, was then
slowly vertically dropped to its position onto the pre-glued
detector tray underneath, whose position could be adjusted with
micrometric resolution at the focus of the camera of a metrology
machine. The DU was kept in its position for more than 24 hours
(the silicon glue polymerization time) by mechanical constraints
and its position measured again afterwards. This procedure was
subsequently repeated for the 4 DUs. The experimental results
obtained with this procedure can be expressed in terms of
parallelism and orthogonality between the innermost edges of the
different DUs, and their absolute distance. The average
parallelism between the 4 units is 15.7 arcsec, with a dispersion
of 10.4 arcsec. The average orthogonality is 9.5 arcsec, with a
dispersion of 9.4 arcsec. The average distance (nominal value
4.000 mm) is 4.014 mm, with a dispersion of 7 $\mu$m. The final
real DP alignment is thus well within the requirement of 60
arcsec.

Further details about the DP integration and alignment procedure,
carried out in collaboration with the MIPOT company, may be found
in Rapisarda et al. (2004). The stability of the alignment, within
the requirements, across the environmental tests (thermal-vacuum
and vibrations) was verified at the engineering model level
(Feroci et al. 2004).

\section{The Collimator and Mask System (CMS)}
\label{coll}

The SA collimator has the double function of limiting the field of
view and be an accurate mechanical support for the coded mask
system. Due to the constraints on trasparency (see \S\ref{intro}),
the collimator structure is in 0.5 mm thick carbon fiber. To the
purpose of shielding the hard X-rays of the diffuse X-ray
background, the collimator walls are covered with tungsten sheets
$\sim$120 $\mu$m thick, that guarantee an absorption greater than
90\% for photon energies lower than 40 keV, in the orthogonal
direction.

The field of view (FOV) was determined as a trade-off between the
main scientific requirements (coverage of the largest fraction of
the GRID observing angle, peak 1-day sensitivity better than 15
mCrab, arcmin-level angular resolution) and the mechanical
(especially volume and mass) and telemetry constraints (25
kbit/s). The chosen solution of a crossed rectangular field,
107$^{\circ} \times$68$^{\circ}$ (FWZR) covers approximately half
of the GRID FOV with an expected sensitivity better than 50 mCrab
(one day integration) in the central
60$^{\circ}\times$60$^{\circ}$ (peaking at 10-15 mCrab on axis),
and with an average telemetry occupation of $\sim$20 kbit/s. (See
\S\ref{sens} for a more extended discussion of the experiment
sensitivity.)

The above FOV is achieved by collimator cells having the same size
of the detector, that is 190 mm x 190 mm, evenly divided in two by
a sect placed orthogonal to the detector strip direction. In this
way the coding direction has a FOV limited to $\pm$53.5$^{\circ}$,
while the non-coding direction extends to $\pm$34$^{\circ}$. The
collimator response is triangular in both directions. The
collimator cell height is 142 mm, and this also fixes the
mask-detector distance, since the coded mask is supported by the
0.5 mm thick carbon fiber top coverage of the collimator itself.
In Figure \ref{fig_coll} the Flight unit of the SA collimator is
shown as seen from the bottom, where the cells are open and
visible.

The radiation blockage of the collimator is guaranteed by the
Tungsten sheets glued to its walls. The sheets placed onto
orthogonal walls could not arrive to touch. The corners between
walls were then covered by a specially manufactured
Tungsten-loaded glue, ensuring approximately the same shielding as
the Tungsten sheets. The interface between the collimator and the
DP was instead covered by a 135$\mu$m thick lead-tape. Finally,
since the carbon fiber and the glue are not very good electrical
conductors, thin copper plates (clearly visible in Figure
\ref{fig_coll}) were glued onto the critical parts of the
collimator walls, ensuring the electrical grounding of the entire
structure.

\subsection{The Physical Design of the Coded Masks}
\label{mask_phys}

The SA experiment has, from the point of view of the coded mask
systems, a \textit{simple configuration} (e.g., in 't Zand 1992),
that is mask and detector have approximately the same size.
Different configurations were excluded by the mechanical and/or
scientific constraints. The two independent units per coordinate
allowed to manufacture the SA mask of the homologous detectors in
a mirror configuration, that is, any part of the sky is seen
through the mask and anti-mask code, one per detector. This
strategy is expected to be highly rewarding when spatial
inhomogeneities are present in one detector. With such a large
FOV, the most common case is the partial occultation of the FOV by
the Earth. In Figure \ref{antimask} a simulated application of the
mask/anti-mask technique shows the advantage of combining the
images of the two homologous detectors when the background is not
uniform (here we show the case of Earth blockage of a fraction of
the field FOV).

The mask/anti-mask technique imposes a mask open fraction of 50\%,
although in background-dominated experiments as SA smaller open
fractions may result in better sensitivity (e.g., in 't Zand et
al. 1994) and, consequently, in a saving in the telemetry
resources. This open fraction also allows to use an Hadamard
sequence for the source code, thus benefiting of an almost
side-lobe free response function for on axis point sources. Our
787-element sequence was thus generated by means of the quadratic
residue algorithm, then conceptually rounded to 788 elements to
match the perfect 50\% open fraction, with the addition of one
element. The size and number of mask elements was driven by the
highest angular resolution matching a minimum factor of 2 detector
over-sampling for each mask element (e.g., in' t Zand 1992 and
references therein). Further details on the mask code selection
and properties for SA, with numerical and Monte Carlo testing of
this and alternative solutions, may be found in Lapshov et al.
(2007).

\subsection{Manufacturing of the Coded Mask System}

At the manufacturing stage, the coded mask of SA, as any other,
deviates from the ideal design under many respects. The major
issue in the manufacturing of the SA coded mask system was doing
242 $\mu$m narrow cuts on single a 440 mm $\times$ 440 mm wide,
117 $\mu$m thick tungsten sheet, satisfying a $\sim$10-15 $\mu$m
accuracy on both the mask elements width, as well as their
position and parallelism with respect to a reference direction,
with the additional requirement of avoiding systematic
displacements in the absolute position of the strips.

Several techniques were tested to comply with the above
requirements. The chemical etching was proven on prototypes to
satisfy the basic requirements and selected for the manufacturing
of the engineering and flight units by the Oerlikon Contraves
Italia, in cooperation with Vaiarelli s.r.l. (Milan). The result
of fabrication is rather satisfactory. A sample of $\sim$400
strips (intentionally manufactured on the same Tungsten sheet used
for the flight unit) shows an average deviation from the nominal
width of 0.2 $\mu$m, with a standard deviation of 8.0 $\mu$m. Very
good results were also obtained with respect to the meniscus along
the mask thickness: the same sample shows deviations within 5-10
$\mu$m from orthogonality. Despite these very good results (a
definite technological challenge considering the wide dimension
and small thickness of the tungsten sheet), a couple of issues
were not fully satisfactory. The deviation of the strip widths was
slightly systematic, resulting in an overall dimension of the mask
approximately 50 $\mu$m wider than the nominal. Although this
implies only a contribution of less than 1 $\mu$m per strip (less
than 0.5\% of the width), it also implies that the outermost
narrow strips are displaced by 1/4 of their width. Another
important point is that the process of gluing the mask onto the
500 $\mu$m thin carbon fiber top layer of the collimator brought
to some arc-like distortions in the linearity of several strips.
The amount of this effect was however limited by the 3
strong-backs supporting the strips every $\sim$4.5 cm. We
developed software techniques to account for both defects when
simulating the source shadowgram during the IROS procedure (see
$\S$ \ref{soft} and Lapshov et al. 2007 for details). The
appearance of the CMS before integration with the DP is as shown
in Figure \ref{cms}.

\section{Mechanical Integration and Alignment}

After the individual "internal" integration of both DP and CMS,
the mask-detector alignment for each DU was thus transposed to the
alignment of the two above systems as rigid bodies. The alignment
was carried out using a 0.1 $\mu$m-precision metrology machine
equipped with an optical camera. A fine-step movement system
allowed to shift the CMS with respect to the DP. Before putting
the CMS in place, reference points on the silicon detectors were
targeted and saved into the memory of the metrology machine. The
CMS was then positioned making its reference points to coincide as
much as possible with the relevant points on the DP.

Since the rotation between the detector and mask strips is the
most critical issue of our optical system, the quality of the
alignment can be evaluated in terms of the angle between these two
items, for each DU. This measurement was carried out, using the
same metrology machine, taking the average strip direction with 8
points per detector strip and 9 points per mask strip. Considering
that the detector is composed of two independent ladders ($\S$
\ref{du}), we studied the misalignment independently for the two
outermost detector strips of each detector, representative of the
two different ladders (the very small misalignment between the two
tiles of a ladder is already averaged by the multiple-point
description of the strip). In Table \ref{tab:alignment} we show
the final results for the two outermost strips of each detector,
as well as their averages. It is worth noticing that these
measurements include all the potential mechanical errors in the
internal alignment of our optical system: the tiles position
within one detector, the misalignment from one detector to
another, and the misalignment of the strips within the coded mask.
In particular, taking into account that the alignment was
basically done between rigid bodies, the larger misalignment in D2
is indicative of an intrinsic misalignment of that specific
detector/mask system.

\begin{table}[h!]
\centering \caption{Measured alignment between the outermost
individual detector strips and their reference mask strips. The
uncertainty in this measurement is only due to the optical
targeting of the strip, and was estimated as $\sim$10$\mu$m
linear, corresponding to $\sim$10 arcsec. }
\bigskip
\label{tab:alignment}

\begin{tabular}{|c|c|c|c|}
  \hline
  Detection Unit   & Average  & Internal  & External  \\
 (coding direction)        & (arcsec) & (arcsec)  & (arcsec)  \\
  \hline
  D0 (Z) & +18.0    & +14.4     & +21.6  \\
  D1 (X) & -18.0    & -18.0     & -18.0  \\
  D2 (Z) & -46.8    & -50.4     & -43.2  \\
  D3 (X) & +14.4    & + 7.2     & +25.2  \\
  \hline
\end{tabular}
\end{table}

Due to the tight schedule of the production of the AGILE Flight
Model, no metrology measurements could be carried out after the
environmental tests (especially vibrations) because this would
have required to disassemble the experiment and qualify it once
again. We then rely on the tests carried out on the SA engineering
model, completely representative of the flight unit from the
mechanical point of view, for which we showed that after the
vibration tests the internal alignment of SA was preserved to
within $\sim$50 arcsec (Feroci et al. 2004).

The above results concern the  alignment internal to the SA
experiment. To the purpose of relating the SA optical axes to the
reference frame of the Star Trackers onboard the AGILE mission, we
positioned a set of 3 optical cubes on the top corners of the
mask. The orthogonality between the faces of the optical cubes is
certified by the manufacturer to better than 3 arcsec. The faces
of these cubes were then measured with the same machine as the SA
optical axes thus allowing to determine the rotation matrix
between the SA Detection Units and the Star Trackers (also
equipped with optical cubes). The alignment between these two
systems was measured after the integration of SA into the AGILE
Payload by means of a system of theodolite and laser tracker
instruments, with an angular uncertainty smaller than 5 arcsec and
an absolute uncertainty smaller than 10 $\mu$m. Only a reference
cube on the AGILE Bus and those on the Star Sensors remained
optically accessible after the satellite integration. Their
alignment was measured again after the satellite environmental
tests (vibration, thermal-vacuum, EMC and acoustic) in order to
verify its stability across odd environmental conditions. In this
set-up only the laser tracker could be used as a measurement
system, not optimal due to the small surfaces involved. The result
of the two measurements shows that the alignment is preserved to
within the uncertainty of the measurement, corresponding to less
than $\sim$2 arcmin.

After its complete integration, the weight of the complete SA
experiment (DP and CMS, excluding SAIEs and harness) was measured
as 5482 g.

\section{The SuperAGILE Interface Electronics (SAIE)}
\label{saie}

The SA Interface Electronics (SAIE) is the digital electronics in
charge of interfacing the SAFEE units with the Payload Data
Handling Unit. SA is equipped with two independent and identical
SAIE boards, each one controlling two SAFEEs, one per coding
direction, providing a full redundancy. A detailed description of
the SAIE functionalities may be found in Pacciani et al. (2007).
Here we recall its main functions:

\begin{itemize}
    \item Regulate power supplies received from the AGILE Power Supply Unit
    for the SAFEE: +2 V and -2 V (from $\pm$2.7 V) independently on each Daisy
    Chain, +5V, -5V and +90V (from 130 V) independently for
    each SAFEE.
    \item Latch-up control, with automatic switch-off, for the +2 V, -5V
    and +5 V lines, by current  monitoring.
    \item Providing digital controls for the SAFEE DACs (for setting the
    XA biases).
    \item Storing and providing configuration strings for XA Reg-In
    registers.
    \item Current-to-voltage conversion of amplitude, strip
    address, group address and multiplicity signal from each Daisy
    Chain.
    \item Assigning 12-bit differential time (2$\mu$s resolution) to
    each
    event. If no event is received within 8.192 ms, a Dummy Event is
    generated to preserve the sequence. The time assignment is
    based on a 1 MHz clock and the GPS pulse per second signal
    received from the PDHU.
    \item Generating 36-bit Absolute Time Events, as reference to
    the event differential time.
    \item Handling simultaneous events on different Daisy Chain.
    \item Applying analog upper thresholds to amplitude signals.
    \item Applying analog upper and lower thresholds to multiplicity signals (MGO).
    \item Handling veto signals from other AGILE subsystems (Silicon Tracker,
    Minicalorimeter, Top and Lateral Anticoincidence).
    \item 12-bit A/D conversion of all signals.
    \item Conditioning temperature sensors (a total of 34 on SA).
    \item Building analog housekeepings (sampling time 16 s):
    voltages, detector bias currents, temperatures.
    \item Building digital ratemeters (integration time 16 s):
    total triggers, veto-rejected counts, energy and
    multiplicity windows-rejected counts, good events.
    \item Generating the electronic calibration procedure:
    a sequence of up to 256 pulses for up to 4 amplitudes for every
    XA channel; the procedure can work automatically for all the 384
    channels in a Daisy Chain, independently for each Daisy Chain.
    \item Generating the threshold scan procedure:
    a sequence of electrical calibrations interleaved with threshold setting
    variations; the procedure can work automatically for all the 384
    channels in a Daisy Chain, independently for each Daisy Chain.
\end{itemize}

The SAIE boards register each event (Good Event, Calibration
Event, Dummy Event or Absolute Time Event) in a 60-bit word,
containing its strip and group address, amplitude, and
differential time. These events are stored in a FIFO (containing
up to 16 events, 1 kbyte) waiting to be read-out by the PDHU.

Each SAIE board is encased in an Aluminum box, 380 mm x 202 mm x
20 mm, for a total weight of $\sim$1.5 kg per SAIE. The power
consumption of each SAIE is $\sim$3.8 W. The 2 SAIE boards are
vertically placed on two sides of the silicon tracker.

\section{The Payload Data Handling Unit}
\label{pdhu}

In addition to the "standard" Data Handling operations, the AGILE
Payload Data Handling Unit (PDHU, based on the 32-bit TEMIC
TSC21020F DSP processor), performs a few but important functions
specific to the SuperAGILE experiment. They can be summarized as
follows:
\begin{itemize}
    \item For Good Events type, converts the 12-bit group and 12-bit strip
    addresses into a
    single 13-bit "digital" address, through a Look-up Table (LUT).
    Four LUTs are stored in the PDHU memory (to account for variations
    due to
    temperature, power supplies, and other environmental effects)
    and their switch is selected by telecommand.
    \item Formats the 60-bit SA events (all types) received by the SAIEs into
    32-bit telemetry events, and creates telemetry packets.
    \item Provides the 12 most significant bits to the event
    arrival time, bringing it from 36-bit to 48-bit (On Board Time,
    based on a reference time received every 1 s from the onboard GPS).
    \item Performs the SA Burst Search and Imaging procedure (see $\S$
    \ref{burst}).
    \item Builds the SA scientific ratemeters: 24 counters (counts every
    0.5 s, in each ladder, in 3 programmable energy ranges).
    \item Generates the automatic switch-off command for the relevant
    SA detector if a latch-up is detected from the SAIE.
    \item Manage the telemetry tables: as a baseline up to 25 kbit/s
    are available to SA (corresponding to roughly 800 counts/s, on
    average over the orbit).
\end{itemize}

The SA-SAIE-PDHU system was designed in order to be able to handle
impulsive SA total rates as high as 100 kHz (corresponding to a
transient source more than 1500 times brighter than the Crab
source on axis) for up to 10 s . This is achieved by the
combination of the SAIE high rate performance and a SA event
buffer (10$^{6}$ events) on the PDHU.

A more general and detailed description of the AGILE PDHU may be
found in Argan et al. 2004.

\subsection{SuperAGILE Burst Search and onboard Imaging}
\label{burst}

Given its field of view and sensitivity, SA has a good chance to
detect and localize cosmic gamma-ray bursts (GRBs), and this is
indeed one of the most important scientific goals of the
experiment, due to its combination with the simultaneous
observation in the GeV energy range. Thus, a particular care was
devoted to this issue during the design of the entire AGILE
payload. In particular, SuperAGILE was equipped with an onboard
trigger logic, able to promptly acknowledge the detection of a
GRB, start an imaging process to localize its coordinates,
activate a fast-communication channel based on the ORBCOMM data
transmission network\footnote{http://www.orbcomm.net} to
distribute the GRB coordinates to any ground- or space-based
observatory almost in real-time.

However, the extremely tight limitations in the onboard processing
capability allocated to this task imposed to make the burst search
and imaging procedure as simple as possible.

\subsubsection{Burst Trigger Logic}

The burst trigger algorithm continuously and simultaneously
compares the count rate of each detector, in two energy ranges, on
6 Short Integration Times (SITs: 1, 16, 64, 256, 1024 and 8192 ms)
with a background rate evaluated over 7 possible Background
Evaluation Times (BETs: 8, 16, 32, 65, 131, 262 s), to be
associated to each SIT by telecommand. Also the type of comparison
can be selected by telecommand between statistical (i.e., the
excess or total counts are compared with the standard deviation of
the background count rate, as evaluated by the average in the SIT)
or static (i.e., the excess counts are compared to a fixed
threshold in counts). The trigger criteria are independently
continuously verified for the 8 ratemeters (4 detectors, 2 energy
ranges) on the 6 SITs. Any possible configuration
"triggered/not-triggered" of this 48-element pattern is then
defined in a look-up table to originate or not a SA Burst Trigger.
Once a SA Burst Trigger is declared in the PDHU, the following
actions take place (when enabled):
\begin{itemize}
   \item the SA imaging procedure is started (see next section);
   \item the download in the telemetry of the event-by-event data for the
   Minicalorimeter (Labanti et al. 2006) is activated over a time span
   dependent on a dedicated algorithm.
\end{itemize}

While processing a SA Burst Trigger, the burst search is inhibited
for a fixed time interval, programmable from ground between 0 and
65535 s.

\subsubsection{Burst Imaging and Coordinate Transmission}
\label{burst_imaging}

The goal of triggering on GRBs is related to the ability to
promptly localize them with arcmin accuracy in the sky. To this
purpose, we devoted cyclic buffers in the PDHU to accumulate
detector images for each of the 4 SA detectors. The integration
time is programmable, typically we expect to set it at
 $\sim$60 s. Two images are accumulated for each
detector at any time: a stored background image and a running
foreground image.

The events used to build the images are the SA Good Events. Due to
the varying attitude of the spacecraft (predicted as
0.1$^{\circ}$/s inside a cone with 1$^{\circ}$ aperture), a
realtime attitude correction is needed in order to recover the
blurring of the images due to the attitude variations. In
practice, in the onboard procedure of building up the images, for
each detected photon instead of incrementing the image pixel
corresponding to the hit strip, another one is incremented whose
coordinate is derived on the basis of the projection of the
attitude error at the time of the photon on that specific
direction. The attitude information is provided in realtime every
0.1 s by the AGILE Star Sensors with arcmin accuracy.

In principle, any region in the sky would require a specific
correction but this implies an iterative procedure that cannot be
performed onboard due to the very limited computing resources
allocable to this task. We thus decided to process all the photons
in the field with an average correction corresponding to the
\textit{a-priori} most probable incoming direction of the burst
(obtained by the analysis of throughput, cm$^{2}\times$ solid
angle), that is 20$^{\circ}$ off-axis. The effect of such an
approximate correction will be a not perfect recovery of the
attitude blurring. Since the high frequency attitude variations
are expected to be symmetric, the net effect for a short timescale
image is expected to be a loss in imaging sensitivity (that is,
the peak in the sky image for a point source will be broader and
thus dimmer) but without significantly affecting the peak position
determination (see also $\S$ \ref{psla}). Long integrations, in
general, are not expected to be used onboard.

When the SA Burst Trigger arrives, the background images of all
the four SA DUs are "frozen", the foreground images reset, and the
accumulation of new foreground (burst) images starts. Indeed, the
background images are always kept $\sim$16-20 seconds "in the
past", in order to avoid to include the early burst photons in the
background images. The end of accumulation of the burst image is
reached when the signal/noise ratio of the image is evaluated to
be large enough to allow a reliable coordinate determination. This
is obtained by means of an algorithm, taking as input the
pre-burst background count rate, the realtime counts and the
desired statistical accuracy. When the stop command arrives, the
realtime accumulation ends, the previously buffered pre-trigger
burst photons are added to the burst image, based on the timescale
(SIT) of the burst trigger.

The next step is the subtraction of the background image from the
image including the burst, after proper re-scaling of the
integration time (so that persistent sources are removed from the
net image), and the deconvolution of the net burst image with the
mask code by means of Fast Fourier Transform. The resulting sky
images are then co-added for the homologous detectors and the
source peak is identified in each of the two coordinates.

A further validation step is foreseen based on the statistical
significance of the peak in the sky images. If unambiguous
coordinates are identified a special burst telemetry packet is
generated. A subset of these information (trigger, SIT and image
duration times, coordinates, pointing, count rates and other
information relevant to the burst) are transmitted to the AGILE
Bus for being sent through the ORBCOMM transponder to the relevant
network for its transmission to the ground as an e-mail message.
The typical delivery time of these messages is anticipated to be
less than 2 minutes for more than 70\% of the messages, being
dependent on the relative instantaneous position of the AGILE and
ORBCOMM satellites.

\section{SuperAGILE Data}
\label{data}

SuperAGILE was designed to transmit photon-by-photon data. The
data type transmitted by the experiment to the AGILE telemetry can
be summarized as follows:
\begin{enumerate}
    \item \textit{Good Event} (GE): each of the two SAIEs drives
    8 daisy chains and it is able to
handle simultaneous events (that is, events on different daisy
chains detected within 2$\mu$s). For each event, SA transmits into
telemetry the event type, its 13-bit strip address (reconstructed
onboard through an address look-up table), 6-bit amplitude, 12-bit
differential time (the time from the previous event, with 2 $\mu$s
resolution);
    \item \textit{Absolute Time} (ABT): the 32-bit Absolute Time events are
generated by each SAIE every 99 events, transmitted as a pair of
events; they are needed to convert the differential time to the On
Board Time (OBT, in turn related to the UTC by means of the
synchronization with the GPS performed every 1 s);
    \item \textit{Dummy Events} (DE): generated by the SAIE
in case no real events are recorded within the maximum range of
the differential time, 8.192 ms; they report the differential time
value;
    \item \textit{Calibration Events} (CAL): the events generated
    during the Electrical Calibration and Threshold Scan
    procedures; these are sequential and not identified with a
    time stamp;
    \item \textit{Analog Housekeepings} (AHK): included in the general AGILE HK
    telemetry packet, they report temperatures, voltages, currents and
    the experiment configuration parameters;
    \item \textit{Digital Housekeepings} (DHK): included in the general AGILE HK
    telemetry packet, they report the 16-s hardware rates of accepted and
    rejected events;
    \item \textit{Scientific Ratemeters}: described in \ref{pdhu}, they
    are included in the general AGILE HK
    telemetry packet;
    \item \textit{Burst Alert}: a telemetry packet generated when an
    onboard SA Burst Trigger is detected and the corresponding function
    is enabled;
    it includes information about the
    event (timescale, counts, coordinates, ...);
    \item \textit{Burst Imaging}: a telemetry packet generated when an
    onboard trigger is detected and the corresponding function is enabled;
    it includes the images integrated
    onboard from which the burst coordinates were derived;
    \item \textit{Imaging Raw}: a telemetry packet generated when the
    event-by-event transmission is inhibited (a solution to apply
    in case of a desired smaller telemetry allocation to SA);
    it includes continuous onboard images
    for each detector with integration time to be set from ground in the range 4 -
    128 seconds.
\end{enumerate}

The data types listed in (1) to (4) are indeed transmitted in the
same telemetry packets, and they need to be separated by the
ground software based on their identification flags.

\section{On-ground Tests and Performance}
\label{perf}

The SA experiment was tested at any step of its integration.
Specific test equipments (TEs) were designed and built to control
and read-out the SAFEE and the SAIE boards (TE-SAFEE and TE-SAIE,
respectively, see Pacciani et al. 2007 for a detailed
description). These equipments allowed to test efficiently the SA
detectors both stand-alone and integrated with the SAIE, by means
of electronic pulses and radioactive sources. When integrated
inside the AGILE payload the tests were performed using the
specific Electrical Ground Support Equipment developed for the
AGILE payload and satellite (e.g., Trifoglio et al. 2004). In this
section we overview the results of the SA functional tests and the
preliminary results of the ground calibrations carried out at
experiment level and after the integration with the spacecraft.

\subsection{Functional Tests Results}

Functional tests were used to optimize the SA performance during
the development phase, and to monitor the SA functionality and
performance across the integration and environmental tests. These
tests can be performed injecting an electrical stimulus
(electrical calibration, threshold scan), or X-ray photons from
radioactive sources or integrating the environmental background.
The electrical calibration is the most used test, because it can
be performed anywhere in a relatively short time, without
requesting any external source, and it tests all the channels
individually. The results of an electrical calibration can be used
to test the noise level, the gain, and the analog response of the
address signals. Since the gain and the noise are drastically
different if the detector strip is connected or not to the ASIC
input channel (e.g., because of a broken wire bonding) this test
can be used as an integrity check, although the detection of
X-rays is the ultimate proof for it. The address signals, in turn,
allow to study the address conversion look-up table as a function
of instrument (e.g., power supplies) or environmental parameters
(e.g., temperature).

\subsubsection{Electrical calibration}
\label{ecal}

Figures \ref{gain} and \ref{fwhm} show the synthetic results of an
electrical calibration of SA obtained in March 2007 during the
latest functional tests just before the shipping of the spacecraft
to the launch base in India. Figure \ref{gain} reports the value
of the gain of each individual strip, in ADC units per fC, as
measured by a linear fit to the response of each channel to 1024
pulses with 4 different amplitudes (here the amplitude corresponds
to 1.35, 1.96, 2.45, 3.06 fC, that is 30.5, 44.3, 55.4 and 69.2
keV nominal). Figure \ref{fwhm}, instead, shows the energy
resolution (Full Width at Half Maximum, in units of keV) derived
for each strip from the same set of data. The reported value is
the average of the values obtained for each of the 4 amplitudes,
converted into keV by using the measured gain individually for
each strip.

The gain is relatively uniform for all the channels, in the range
200-300 ADC/fC. Also the energy resolution is around 8 keV for
most channels. Only the detection unit D2 has a significant
fraction of channels with a lower gain and a higher level of
noise. For D2, in those ASICs interfacing the edges of the two
ladders, this is due to an accidental damage occurred during the
ground calibrations, likely due to an electrostatic discharge.
This event caused approximately 200 channels, distributed over 4
chips, to be permanently disabled, while the others channels of
this unit survived but some of them with a higher level of noise.
In fact, the same event provoked also a permanent increase in the
dark current of the relevant detector, now $\sim$3.5 $\mu$A,
compared to the typical $\sim$1 $\mu$A of the other three
detectors, at room temperature. From these plots it is also
possible to identify the XA channels not connected to the
corresponding detector strip because they show a higher gain and a
better energy resolution, due to their smaller capacitive load.

\subsubsection{Threshold Scan}
\label{threshold}

This type of test is aimed to establish the relation between the
discriminator threshold voltage (in mV) and the charge (in fC) for
each channel of each XA ASIC chip. It consists of a sequence of
electrical calibrations, scanning different values of the
discriminator threshold. An operational definition of threshold is
given as the discriminator reference voltage for which for a given
charge injected to the test input of the ASIC the number of counts
detected corresponds to half of the number of injected pulses. The
outcome of this procedure thus allows to obtain an energy
calibration of the setting of the analog thresholds, for a given
experimental set-up. This procedure is significantly time
consuming, as it consists of an electrical calibration repeated as
many times as the number of threshold steps, typically 25. It also
produces a huge amount of telemetry: with 4 amplitudes, 128 pulses
per amplitude and 25 threshold steps it requires $\sim$1.8 Gbit of
telemetry, corresponding to $\sim$12 orbits of standard operation.
This implies that this procedure will need to be considered for
use in orbit only in exceptional cases, at least in its complete
form.

In Figure \ref{thr} we show the threshold of each channel, in keV,
for a given value of threshold DAC setting, that is the same
nominal voltage reference.  The commandability allows to set
different voltage settings for each daisy chain (with a step
corresponding to $\sim$1.5 keV). To decrease the dispersion
between channels belonging to the same daisy chain, a 3-bit DAC
setting is available for each of the 384 channels, and can be set
through its shift register (Reg-In). The threshold equalization
using these internal DACs is extremely complex and cumbersome,
because they induce correlated variations to all channels of the
daisy chain. In addition, the threshold setting is significantly
temperature dependent. The analog threshold equalization was
devoted a specific one-week test run that allowed to reach a
threshold configuration dispersion of $\sim$0.8 keV (rms, over the
entire SA experiment), compared to the intrinsic $\sim$2.5 keV for
a non equalized set of thresholds. The technique and procedure
used, as well as the details of the results, will be extensively
discussed elsewhere (L. Pacciani et al., 2007b). The analog
threshold configuration shown in Figure \ref{thr} is the one
loaded into the SA front-end electronics at the time of the
launch.

\subsubsection{Environmental Background}

The electrical calibration provides a measurement of the
electronic noise through the measurement of the energy resolution
in response to mono-energetic pulses. However, the energy
resolution is only indicative for the discrimination threshold,
that requires dedicated measurements. In particular, setting the
experiment in photon acquisition mode, with a reasonable threshold
setting, provides an energy spectrum whose components are expected
to be only the environmental background and the high-amplitude
tail of the electronic noise. The former can be easily
distinguished by its flat spectrum, thus allowing to estimate the
count rate of the electronic noise (expected to peak at low
amplitudes) as a function of the threshold voltage. Fig. \ref{bkg}
shows a representative energy spectrum of the background, obtained
with unit D3 and a nominal threshold of 17 keV, with the fine
thresholds setting in its flight configuration. The low energy
peak due to the electronic noise and the flat environmental
background components are clearly distinguished. The broad peak at
$\sim$95 keV is due to the saturations of the individual channels,
broadened by the gain equalization of the $\sim$1500 active
detector strips. The setting of the analog thresholds in orbit
will be defined on the basis of the detected level of the
electronic noise (highly dependent on the temperature of the front
end electronics) and the allowed telemetry load.

\subsubsection{Radioactive X-ray sources}
\label{test_x}

 The Electrical Calibration procedure allows to calibrate the
response of the signal processing chain, from the charge
preamplifiers inside the ASICs to the digital processing of the
event. However, the photon detection capabilities need to be
tested using X-ray photons. This allows to avoid the uncertainties
about the test capacitor in the ASIC, that converts the test
voltage pulse into a test charge, and the noise of the test
circuitry itself (it has to be noted that the charge equivalent to
10 keV in SA is obtained with a test pulse as low as 0.88 mV). In
Fig. \ref{cd109} and Fig. \ref{I125} we show the energy spectrum
of the Cd$^{109}$ (main line at 22 keV) and I$^{125}$ (line
complex at 27-32 keV) radioactive sources obtained with the D0 and
D1 units, respectively. From such type of measurements the
linearity curve for each of the four SA detection units can be
obtained. Indeed, being the analog chains all independent, one
should derive a linearity curve for each channel, but this
conflicts with the counting statistics that can be achieved with
reasonable experimental set-ups and integration times. As shown in
Del Monte et al. (2007) the channel-energy conversion obtained
with X-ray photons test is linear up to 120 keV, well outside the
nominal SA energy range. The value of the slope coefficient is 3\%
larger than the corresponding one obtained with the electrical
calibration. This is most likely due to the average difference
between the real test capacitors of the ASIC channels with respect
to their nominal 0.5 pF value.

From the test runs with radioactive X-ray sources we can also
derive an estimate of the energy resolution, independent from the
electrical test pulse properties. For example, for the lines of
the I$^{125}$ complex, dominated by the $\sim$27 keV line, we
derive a value of (8.3$\pm0.1$) keV, (8.3$\pm0.1$) keV,
(8.4$\pm0.1$) keV, and (8.2$\pm0.1$) keV FWHM for the 4 SA
detection units, respectively. The consistency of these values
with those obtained with the electrical calibration procedure
($\S$ \ref{ecal}) implies that, with respect to its noise
performance, the latter measurement is not significantly affected
by the layout of the onboard test circuit, including its section
internal to the ASIC itself, and can be taken as a good indicator
of the instrument performance in orbit, where monochromatic lines
from radioactive sources are no longer available.

\section{On-ground Calibrations}

The SA experiment was calibrated on ground in three subsequent
steps: at detection plane level (before the integration with the
mask and collimator system and with the SAIEs) in IASF Rome on
March 2005; at experiment level (with the experiment fully
integrated but before the integration with the AGILE payload) in
IASF Rome on August 2005; at satellite level at the Carlo Gavazzi
Space facilities in Tortona (Alessandria, Italy) on January 2007.
The details of the 2005 calibrations at subsystem level are given
in Evangelista et al. 2006, and Donnarumma et al. 2006. An
extensive and detailed analysis of the SA performance on ground,
as derived by the final ground calibrations, will be published in
a forthcoming paper. Here we summarize the main results.

\subsection{Calibrations at Experiment Level}

\subsubsection{Calibration of Detection Units}

The detection plane was studied on March 2005 by means of standard
radioactive sources (at 22, 32, 59 and 122 keV), and a
quasi-parallel beam obtained by collimating an X-ray tube to a
rectangular beam with the short side narrower than the detector
strip width (approximately 100 $\mu$m by 1.0 mm/1.8 mm at the
Silicon) and with a divergence of approximately 0.7 arcmin. The
spectrum was a bremsstrahalung continuum, peaking around 35 keV.
Using these two set-up systems we studied the detector linearity,
energy response, the dead time and the uniformity of the detection
area. At the time of these measurements the SAIE boards were not
fully completed and tested yet, thus the tests described in this
section were carried out directly interfacing the SAFEE boards to
their Test Equipment (see Pacciani et al. 2007 for details).

The linearity and energy resolution issues were discussed in $\S$
\ref{test_x} and in the cited reference papers. The study of the
detection uniformity showed a loss of about 4\% of the counts at
the border region between pairs of adjacent strips, likely caused
by charge loss due to border effects of the electric field.
Similarly, we found that the edge strips of the tiles have a
larger efficiency, due to the fact that the guard ring, here
limiting the strip, is farer than the strip pitch elsewhere in the
detector. The increase in active area of the edge strips of each
Si tile is $\sim$10\%, but in this case it is almost negligible,
due to the fact that it applies to only 4 strip per detector. In
any case, the above effects will be included in the instrument
response matrix.

The instrument response to high count rates was tested with the
X-ray beam, sending up to 1500 counts/s on each strip,
individually. For one single strip this count rate corresponds to
a source more than 10$^{6}$ brighter than the Crab Nebula.
Considering one daisy chain, instead, such a rate corresponds to
$\sim$400 times the Crab. Despite such a high rate, the detector
response showed no distortions and the power consumption did not
change appreciably. The estimated dead time per daisy chain was
consistent with that derived from the timing as set by the
relevant parameters (TrigDelBias, TrigWBias, ResWBias) for that
acquisition, approximately 240 $\mu$s. In the flight configuration
set-up this value (the dead time per daisy chain) was lowered down
to 80 $\mu$s (see $\S$ \ref{safee}).

Further details on the set-up and the measurements may be found in
Evangelista et al. (2006).

\subsubsection{Imaging Response at Experiment level: parallel beam}

The same collimated X-ray tube apparatus was used to test the
imaging response of the experiment, after its definitive
integration and alignment. The set-up included the SAIE boards and
the measurements were therefore carried out using the SAIE Test
Equipment (see Pacciani et al. 2007 for details).

The goal of this measurement was to obtain information about the
point spread function (PSF) of the experiment, thanks the small
divergence of the impinging beam thus providing a good
approximation of a parallel beam. However, a recognized
significant limitation to this measurement is related to the small
extension ($\sim$ 1.8 mm) of the beam along the mask strip. This
implies that the real mask is sampled only in that specific
section, and any local defect will have a much larger impact on
the response than it will be in reality, when the complete mask
will be sampled and the imperfections averaged out. Since a
complete scan of the mask surface was made impossible by the
schedule constraints, this measurement was complemented by a
calibration of the experiment imaging response to a global
illumination by omnidirectional sources (see next section),
although in this case the beam divergence must be corrected for
and the source extension subtracted. Again due to time
constraints, the calibration with the X-ray tube was carried out
on axis on the four detectors, and at 10$^{\circ}$ and
20$^{\circ}$ off axis for one detector only.

In Fig. \ref{tube_onaxis_zoom} we show the deconvolved images of
the (approximately) on-axis beam, as observed by the four
detectors in the energy range 27-45 keV. This energy selection was
applied because at the time of this measurement (August 2005) the
analog threshold equalization had not been carried out yet (see
$\S$ \ref{threshold}) and the non-uniformity of efficiency among
the different channels, due to the their intrinsic analog
threshold dispersion, would have artificially affected the image
reconstruction.

The 4 images show the source counts mostly distributed over three
sky bins, as expected by the design of SuperAGILE. Indeed, the
distribution of counts over the reconstructed sky pixels is a
complex issue (see Evangelista et al. 2006 for an extensive
discussion). Due to the discrete pixels of the detector, for an
ideal mask and detector, the counts to be expected in the three
pixels on the basis of the ideal PSF depend on the \textit{exact}
angular position of the source, that is the relative phase between
the mask element and the detector element, with a period of one
detector pixel (3 arcmin). At phase 0, the counts are distributed
according to the proportion 0.5:1.0:0.5 (meaning that the sum over
the 3 bins is twice the source counts). In the "worst case", they
will be distributed over 4 pixels, according to the ratio
0.25:0.75:0.75:0.25. Intermediate phases provide intermediate
distributions of counts over 3 or 4 pixels. Unfortunately, our
set-up did not allow a measurement of the beam inclination
accurate enough, thus the phase remains an unknown for all the
measurements, different for each one, being the Detection Plane
rotated at every measurement and having each of the four
detectors, at this level of accuracy, its own independent optical
axis, related to the fine details of its construction. Since the
relative phase can only result in the distortion of the symmetry
of counts distribution, the almost perfect symmetry of image from
D2 and D3 implies that the beam was very close to phase 0 in these
cases, and this allows to confirm that for these two systems
(detector plus mask) the response is very close to the ideal one.
The most symmetric case, D3, shows significant counts over 5
contiguous pixels, with intensity with respect to the total source
counts according to 0.04:0.50:0.94:0.49:0.04. The difference
between this proportion and the ideal one is understood as due to
the physical effects (e.g., mask transparency at high energies,
finite thickness of the mask, Compton scattering of photons on the
mask, ...) and to the distortions of the specific section of the
real mask sampled by this measurement. A deeper analysis of these
calibration data is under way in order to understand in greater
details the instrument imaging response function. The results will
be reported in a future publication.

The same measurement was carried out at two off-axis angles for
unit D0. In Fig. \ref{tube_offaxis_zoom} we show the relevant
images together with the on-axis measurement on the same unit, for
comparison. These plots allow to appreciate how the SA PSF is only
weakly dependent on the source position in the FOV, on at least a
large portion of its effective 2x1D FOV
($\sim$60$^{\circ}\times$60$^{\circ}$, at zero response).

Indeed, some distortions in the images are expected when the
source is off-axis, increasing in amplitude as the off-axis angle
increases. This is due to the incomplete sampling of the mask
code. In fact, our selected mask code is fully balanced only for
on-axis sources, while the rest of the field is partially coded
and the images show the effect of the coding noise. In Figure
\ref{tube_offaxis_full} we show these distortions in the above
measurements by showing the SA image of the full sky, in place of
the zoom-in on the source. As discussed in $\S$ \ref{mask_phys},
these distortions are largely compensated when the homologous DUs
are used to image the same source on the same coordinate, thanks
to the mask/anti-mask imaging technique. Alternatively, balancing
techniques can be applied also for individual DU detections. On
the other hand, the same figure allows to appreciate the nice
flat-lobe response of the SA optical system to on-axis sources. At
large off-axis angles a contribution to the peak broadening is
expected also as an effect of the inclined penetration of photons
in the silicon.

\subsubsection{Imaging Response at Experiment level: divergent beam}
\label{finite1}

 As discussed in the previous section, due to
set-up and time limitations, the parallel beam illumination of SA
could provide the experiment imaging response only for a small
($\sim$2 mm) section of the mask. To the purpose of calibrating
the average instrument response, the only reasonable alternative
was to use omnidirectional radioactive X-ray sources. To this
purpose we developed a procedure (see Donnarumma et al. 2006 for
details) to correct for the beam divergence and the physical
source extension. With such procedure, we were able to derive
properties of the SA response to an infinite distance source using
finite distance instead. Of course, the procedure is increasingly
accurate as the source distance increases. On the other hand, this
leads to larger integration times, and a trade-off needs to be
found. The final aim of these measurements at experiment level was
to test and calibrate the procedure in view of the final
experiment calibrations (see next section), as well as deriving
some early information about some experiment parameters and the
imaging response. In fact, during the August 2005 calibration
campaign only a short time could be allocated, and the set of
measurements is relatively poor thereof.

The distance between the source and the Detection Plane was set at
$\sim$200 cm, leading to a beam divergence of $\sim$6$^{\circ}$
within one detector. The measurements were carried out with a
Cd$^{109}$ radioactive source (main line at 22 keV) at
0$^{\circ}$, 10$^{\circ}$, 20$^{\circ}$ and 30$^{\circ}$ off-axis
for D0, and with a Cs$^{137}$ source (line of interest, 32 keV) on
axis only. The exact position of the radioactive sources with
respect to SA was measured by means of a laser tracker, having
mounted optical reference targets on the source holder and optical
cubes on SA. Since the reconstruction procedure for
finite-distance source imaging is sensitive to the ratio between
the mask-detector distance and the source-mask distance, the
optimization of the image reconstruction allowed to derive from
multiple measurements an independent estimate of the mask-detector
distance. The derived value of (142.44$\pm$0.10) mm for unit D0
(for which the full set of measurements is available) is
consistent  with both the experiment mechanical design and the
measurement derived by the detailed micrometric precision mapping
of the experiment carried out with the metrology machine during
the experiment integration.

In Figure \ref{cal_aug2005} we show a sample of the reconstructed
source images obtained with D0 during this calibration session.
(For a more complete and detailed description we refer the reader
to Donnarumma et al. 2006). Although these images are not yet
corrected for the source extension (approximately 8 arcmin) they
confirm the expected PSF up to 30$^{\circ}$ off-axis. The detailed
determination of the PSF parameters from these measurements
requires an accurate mapping of the space distribution of the
radionuclide, partly responsible for the distribution of counts.
This measurement is currently planned but not yet carried out.

\section{Ground Calibrations at Satellite Level}
\label{finite2}

The final SA ground calibrations were carried out for one week on
January 2007, when the experiment was onboard the fully integrated
satellite. The set-up was based on the experience of the August
2005 calibrations, with radioactive X-ray sources located at
$\sim$200 cm from the SA mask. The source support was studied and
manufactured in order to avoid some systematic uncertainties on
the radionuclide location revealed during the 2005 calibrations,
and it was mounted on a movable structure, guaranteeing at the
same time the satellite safety, the source mobility at a distance
of $\sim$4 m from ground (the satellite was in vertical position,
placing SA $\sim$2 m above the floor) and the required micrometric
stability. The micrometric position of the source was targeted
with a laser tracker also in this campaign, with respect to
reference points on the spacecraft structure, and to the reference
optical cubes aligned to the AGILE star sensors.

For this calibration campaign 3 radioactive sources were used:
Cd$^{109}$ (main line at 22 keV), I$^{125}$ (line complex at 27-32
keV) and Am$^{241}$ (main line at 59.5 keV, plus line complex at
14-26 keV). The largest sampling was carried out using the
brightest source, Cd$^{109}$. In order to save time, the same
measurement was used for the 4 DUs, implying that slightly
different off-axis positions were sampled for each DUs. In
particular, we took as a reference D0 and built a "cross" in its
field of view, from -50$^{\circ}$ to +50$^{\circ}$ off-axis in
both coding and non-coding directions, step 10$^{\circ}$. The
central square included in -20$^{\circ}$ - +20$^{\circ}$ in both
directions was also filled with 10$^{\circ}$ steps. Finally, we
took measurements at (-30$^{\circ}$,-30$^{\circ}$),
(-30$^{\circ}$,+30$^{\circ}$), (+30$^{\circ}$,-30$^{\circ}$) and
(+30$^{\circ}$,+30$^{\circ}$), for a total of 55 FOV positions
sampled with the Cd$^{109}$. The other two sources were instead
used to sample fewer position near the center of the FOV, for a
total of 12 measurements with the I$^{125}$ source and 3
measurements with the very weak Am$^{241}$, for which the distance
had to be decreased down to $\sim$100 cm. As discussed above, for
the other 3 SA DUs the off-axis angles were slightly different,
due to the finite distance of the source, the relative off-set
being in the range of 5$^{\circ}$-10$^{\circ}$, depending on the
source position.

In Figure \ref{cal_jan2007} we show a sample of the zoomed source
images for one of the SA DUs. The analysis of the complete data
set collected during these recent calibrations is still under way.
A detailed report will be given in a future publication. Here we
remind the goals we aim to, for some of which preliminary results
are already being obtained, for each of the 4 DUs: determination
of the PSF as a function of energy and position in the FOV (this
will require also the subtraction of the source extension, after
its measurement), calibration of the relation between the detector
pixel and the exact off-axis angle, response as a function of
energy, map of obscuring structures in the FOV (mostly the support
structure of the anticoincidence system and some harness),
estimation of the accuracy of localizing single point sources.

During the same calibration campaign some time was devoted to
study the effect of a source location outside the FOV, i.e.,
searching for leaking directions in the experiment shielding to
the diffuse X-ray background. A few leaking or partially leaking
directions were found, but none of them seriously affecting the
experiment performance. They are being accounted for in the
experiment response function.

\section{The Ground Software: SuperAGILE Data Analysis System (SADAS)}
\label{soft}

Despite the very limited resources, SA is a photon-by-photon
experiment and requires a full data processing on ground. The SA
Scientific Software was entirely developed at IASF Rome and it is
planned to run autonomously at the AGILE Data Center (ADC) located
at the ASI Science Data Center (ASDC) in Frascati for the standard
analysis and at IASF Rome for the refined analysis, with human
support. The source position and fluxes will be made publicly
available by ASDC in the shortest time to the whole science
community through the a dedicated SA web page.

The SA data processing starts with the AGILE data transmission
from the spacecraft to the Malindi ground station. From there,
data are transmitted to the ASDC where the telemetry is
pre-processed and archived. The pre-processing is carried out
following the DISCOS system already used on previous astronomy
missions (e.g., Gianotti et al. 2001) and transforms the raw
telemetry data (Level 0) to a FITS standard format (Level 1). The
subsequent processing is in charge to the SA Pipeline, that will
run automatically at every ground contact (i.e., nominally every
100 minutes). The basic conceptual steps of the pipeline are:

\begin{enumerate}
    \item Data Quality Filter;
    \item Separation of data types: events, calibration, absolute
    time, dummy;
    \item Creation of photon list;
    \item Attitude correction;
    \item Source extraction by an Iterative Removal of Sources procedure;
    \item Source identification.
\end{enumerate}

Each of these steps is a complex procedure itself. We refer to
Lazzarotto et al. (2007) for a detailed description of the SA
ground software and pipeline processing. At different steps of
pipeline, organized in 4 stages, data products will be archived.
The final data products, in standard FITS formats, concerning the
position and identification of the detected sources and their flux
level will be transferred to the ASDC public archive for use by
the general community.

The SA data processing will be carried at different temporal
stages. At any contact (orbit) a prompt analysis of the data
collected during the last available orbit will be carried out.
This is aimed to obtain the fastest reaction to transient events.
After the orbital processing is completed, the same data are added
to those of the previous orbits looking at the same field, in
order to identify weaker sources and improve the localization of
unknown sources. Finally, the same data will then be reprocessed
with the final auxiliary data (e.g., reconstructed attitude,
ephemeris, ...) to provide their final scientific products.

\subsection{Real-time Alerts of X-ray transients}
\label{alerts}

As discussed in $\S$ \ref{burst}, part of the onboard software was
designed to trigger and autonomously localize fast hard X-ray
transients, namely gamma-ray bursts and objects with similar
phenomenology. The coordinates are sent in quasi-realtime to
ground through the ORBCOMM satellite link.

The SA ground segment thus includes a section that is devoted to
process the ORBCOMM messages at their arrival, and to perform the
refined data analysis after the satellite ground contact has made
the relevant SA data available to the ASDC. The main features of
this section are: validating the ORBCOMM message, validating the
onboard trigger, confirming the reliability of the onboard
localization, assign localization uncertainty, verify the
coordinates against source catalogs, disseminate the coordinates
of the hard X-ray transient to the general community through the
standard communication channels (e.g., BACODINE, GCN, ATel, ...).

After the early processing, based on the data transmitted through
the realtime message, the same trigger is validated and the
trigger properties are refined by an on ground processing of the
complete SA data set. In this phase the data of the other AGILE
subsystems (GRID and MCAL) will be available as well, allowing to
determine the high energy properties of the event in the $\sim$MeV
and hundreds of MeV ranges. The results of these off-line analysis
will also be distributed as quickly as possible to the widest
community.

\section{Astrophysical Expectations}
\label{astro}

SuperAGILE was designed as the hard X-ray monitor of the AGILE
gamma ray imager, and needed to fit tight design constraints.
However, considering the emission properties of the known
gamma-ray sources and the SA expected sensitivity, the classes of
known sources where the two instruments are really expected to
work together are only a few (but SA is in the best position to
extend this set to other classes of sources, especially variable
ones). However, SA will also work as an autonomous instrument,
taking advantage of its wide field of view and reasonable
sensitivity. In this section we briefly review the expected
performance of the experiment and outline the potential
contribution of SA to the current scenario of the observational
X-ray and gamma-ray astronomy.

\subsection{Effective Area and Sensitivity}
\label{sens}

The SuperAGILE response was computed by means of analytical
calculations and Monte Carlo simulations, based on a model
describing the geometry and mass of the experiment in great
detail. In Fig. \ref{area_vs_ene} we show the effective area, as a
function of energy and angular position of the source, as computed
by means of analytical calculations. The peak of the effective
area is at about 14 keV, just below the effective SA lower energy
threshold (determined mostly by the electronic noise), and it is
due to the absorption of the intervening material (mainly the
plastic anticoincidence and its support structure) at low energy,
and the Silicon efficiency decrease at high energy. Due to the
inclined penetration of radiation in Silicon, at higher energies
the peak shifts as the off-axis angle increases, reaching $\sim$16
keV at 50$^{\circ}$.

Assuming a source with an energy spectrum similar to the Crab
Nebula ($I(E)\sim E^{-2}$) and simulating the SA response to the
source and to the diffuse X-ray background, we can compute the
expected sensitivity to a single source as a function of its
location inside the field of view. With an integration time of
50,000 s (effectively, about one day of observation, for typical
orbital constraints in the AGILE low-earth orbit) the results of
our simulations are shown in Fig. \ref{sens_vs_fov}, in terms of
sensitivity for two homologous detectors, in the energy range
15-45 keV, in units of mCrab (1 mCrab corresponds to 4$\times
10^{-4}$ photon cm$^{-2}$ s$^{-1}$ or 1.5 $\times$10$^{-11}$ erg
cm$^{-2}$ s$^{-1}$ in this energy range), as well as for the
combination of the 4 detectors, imposing an independent detection
on each coordinate. The SA sensitivity thus appears rather "flat"
over the central $\sim$60$^{\circ}\times$60$^{\circ}$ of the field
of view, going from $\sim$15 mCrab in the central region to
$\sim$50 mCrab at $\sim$30$^{\circ}$ off axis, making the
simultaneous monitoring of large regions of the sky an achievable
goal for the experiment.

\subsection{Point Source Location Accuracy}
\label{psla}

The statistical point source location accuracy (PSLA) in coded
mask experiments like SA depends on the ratio between the position
resolution of the detector
 and the factor $f \times SNR$, where $f$ is
the mask-detector distance and $SNR$ is the signal-to-noise ratio
of the specific measurement (e.g., in 't Zand 1992). Given the SA
3 arcmin on-axis pixel size, the PSLA for a 5$\sigma$ source
should be already in the sub-arcmin region. At off-axis angles the
SA pixel decreases in angular size (down to 1.5 arcmin at the edge
of the FOV), but the effective area is far smaller.

We verified the above relation by taking a long on-axis
measurement with the 22 keV source collected during the ground
calibration and dividing it in a large number of statistically
independent measurements, with SNR$\sim$5 and $\sim$10. In Figure
\ref{fig_psla} we show the distribution of the barycenter of the
reconstructed image in one SA DU. Despite the measurement suffered
the limitations of our finite-distance calibrations (see $\S$
\ref{finite1} and $\S$ \ref{finite2}), the measured distributions
show an $rms$ of 0.8 and 0.4 arcmin, respectively, only slightly
worse than the theoretical 0.6 and 0.3 arcmin limits.

The imaging procedure of SA in its flight operation includes the
attitude correction, needed to recover from the large attitude
variations expected in orbit, due to the bus attitude control
system. In fact, the AGILE spacecraft is only able to stabilize
the attitude within 1$^{\circ}$ of the requested target, with
variation as large as 0.1$^{\circ}$/s. In addition to this large
attitude variation on short timescale, the AGILE bus requires the
solar panels to be  orthogonal to the Sun within 1$^{\circ}$,
implying a continuous drift of the pointing to keep this
constraint satisfied. The star trackers will provide an attitude
reconstruction with an accuracy of $\sim$1 arcmin with a 10 Hz
frequency. Thus the position of all the SA photons must be
corrected to their original position, with the strong limitation
of the one dimensional imaging and the discrete pixels with size
corresponding to 3 arcminutes. Indeed, the SA PSLA will be most
likely effectively limited by the systematics in the satellite
attitude stability and reconstruction.  In Fig. \ref{att_recon} we
show an example of the results of the attitude reconstruction
procedure developed by the SA team, applied to a simulated image
blurred by the expected in-flight attitude variation. The result
of the correction is expected to preserve the "ideal" (that is,
with no attitude variation) signal-to-noise ratio to within 95\%.

What above concerns the on-ground processing of the SA data.
However, as discussed in $\S$ \ref{burst_imaging}, the onboard
burst triggering and localization also requires a real-time
attitude correction of the events. Since this correction is
dependent on the source location in the FOV, it requires an
iterative procedure that cannot be supported by the onboard
computing power. As a consequence, an approximate correction is
performed, using an average correction parameter that is exact
only for $\sim$20$^{\circ}$, the off-axis angle at which the
experiment has its largest throughput. For any other source
position in the FOV, the attitude correction will be non optimal,
resulting in a widening of the source peak, and a consequent loss
in the source image significance. The fourth panel in Fig.
\ref{att_recon} shows the effect of such an approximation. This
will affect the onboard localizations of gamma-ray bursts
distributed in real time through the ORBCOMM link (see Sect.
\ref{pdhu}). A ground analysis of the same data, to be performed
after the data downlink at the passage over the Malindi Ground
Station will be able to optimize the attitude correction, and the
data analysis in general, and distribute refined and more accurate
coordinates of the detected transient. This will happen with a
delay varying between few minutes and few hours, depending on the
transient occurrence time with respect to the data download
carried out every $\sim$100 minutes at the passage over the ground
station.

\subsection{The Expected SuperAGILE View of the X-ray and
Gamma-ray Sky }

The main SA science goal relates to the simultaneous monitoring of
the gamma-ray sky observed by the AGILE primary instrument, the
gamma ray imager. The classes of astrophysical sources known to
emit in the energy band 30 MeV - 50 GeV to a level of intensity
compatible with the AGILE sensitivity can be listed as: gamma-ray
pulsars, supernova remnants, Active Galactic Nuclei (mostly
Blazars), gamma-ray bursts and the unidentified EGRET sources. The
known hard X-ray emission of the gamma-ray pulsars and supernova
remnants is largely below the SA sensitivity level, except for the
case of the Crab Nebula and Pulsar, and possibly the Vela Pulsar.
The X-ray flux of blazars is typically near or below the
sensitivity threshold of SA, but in some cases outbursts of these
sources brings an hard X-ray emission detectable by SA (e.g., Mkn
501, Pian et al. 1998). Also the brightest members of other
classes of AGNs (quasars, Seyfert 1 and 2 galaxies, radio
galaxies), are expected to be detectable with SA. In a few cases
these are known emitters of gamma-rays, as 3C 273 and 3C 279
(e.g., Hartman et al. 1999).

Concerning the unidentified EGRET sources, almost nothing is known
about their nature. The quest for hard X-ray counterparts has been
recently carried out by INTEGRAL in an energy range similar to the
SA one. No obvious counterpart has been found, except possibly in
a few cases (e.g., Di Cocco et al. 2004, Foschini et al. 2005).
Therefore, several or most of the unidentified EGRET sources are
unlikely to be persistent and bright hard X-ray sources. But they
may well be transient emitters and the very large field of view of
SA might become crucial in their observation, in this respect.

The situation becomes far more favorable as it comes to the
gamma-ray burst sources. The prompt emission of these objects is
largely above the SA sensitivity limit, even on their typical time
scales (from tenths to tens of seconds). Based on its FOV and
sensitivity, SA is expected to detect and localize gamma-ray
bursts at a rate of $\sim$1-2 events/month. This expectation was
the driver to equip the system with an autonomous triggering and
localization system, and with the fast down-link of the event
coordinates. Although the SA localizations will  be just a few
with respect to those currently obtained by the Swift mission
(Gehrels et al. 2004), they will be uniquely characterized by the
simultaneous observation in gamma-rays. The AGILE data combined
with the multi-wavelength follow-up studies might then be able to
provide a complete view of GeV-emitting gamma ray bursts,
including their distance, energetics and counterpart.

The SA operation as hard X-ray monitor of the AGILE gamma-ray
imager is complemented by its "independent life", as a wide field
X-ray experiment. The large field of view, high angular resolution
and moderate sensitivity makes SA suitable for (quasi-)all-sky
monitoring. The hard X-ray sky is highly variable and new sources
are continuously discovered, especially in the class of compact
objects. Monitoring of known sources provides crucial information
for their long-term behavior and allows observations with
sensitive telescopes to be carried out at proper times and source
states. The most favorable candidates for SA are the galactic
sources, and this is also favored by the expected AGILE pointing
plan, foreseeing large part of the observing time to be spent on
the Galactic Plane. As an example, in Fig. \ref{gal_center} we
show a simulation of a one-day SA observation of the Galactic
Center region (sources as described by the second INTEGRAL ISGRI
catalog, Bird et al. 2006). The Monte Carlo simulation, including
a full description of the SA geometry, takes into account the
simultaneous contribution of all sources and of the diffuse X-ray
background. The data processing of this simulation with the SA
ground software includes the iterative removal of sources, IROS.

\section{Summary and Conclusions}
\label{concl}

In this paper we described the design, manufacturing, integration
and ground performance of the SuperAGILE experiment, the hard
X-ray monitor of the AGILE gamma-ray space mission. The ground
calibrations demonstrated the nominal performance of the
experiment in the 18-45 keV energy range. Lower energy thresholds
may be achieved in orbit if the experiment temperature will be
lower than the one at which it was operated during tests and
calibration (between 35 and 40 degrees at the front-end). The
measurement of the point spread function confirmed the expected
three-bin triangular shape, with a 6 arcmin full width at half
maximum, and a point source location accuracy that will likely be
limited to $\sim$1 arcmin by the spacecraft attitude
reconstruction. The on-axis sensitivity is expected in the range
of $\sim$15 mCrab for a one-day exposure on an extragalactic
field.

SuperAGILE was launched on 23$^{rd}$ April 2007 in an equatorial
orbit (540 km altitude, 2.4$^{\circ}$ inclination). The initial
operations in orbit showed a nominal performance. The check-out
phase will continue over the first two months and will be followed
by a two-month science verification phase. The in flight
performance and observational results will be subject of future
papers.

\section{Acknowledgments}
\label{ackno}

AGILE is a mission of the Italian Space Agency, with
co-participation of INAF (Istituto Nazionale di Astrofisica) and
INFN (Istituto Nazionale di Fisica Nucleare).

 Along the ten years of the SuperAGILE design,
development, manufacturing, integration and testing a number of
people contributed from the involved companies of from research
institutes to this work at any level, administrative, technical or
scientific. We are pleased to acknowledge their individual
contributions, without which the SuperAGILE project would hardly
have been possible: the AGILE Co-Principal Investigator G.
Barbiellini (INFN - Sez. Trieste), L. Barbanera, A. Bartocci, G.
De Paris, C. Pala, G. Sabatino, B. Schena, O. Uberti (INAF/IASF,
Rome), A. Morbidini (INAF/IFSI, Rome), L. Bettinali, A. Lo Bue, P.
Rossi, L. Semeraro (ENEA, Frascati), A. Generosi, V.
Rossi-Albertini, E. Verona (CNR, Rome-Tor Vergata), A. Russo (CNR,
Naples), M. Conte, G. De Paris, L. Soli, A. Zambra, A. Trois
(current and past AGILE System Team), S. Mereghetti, S. Vercellone
(INAF/IASF, Milan), A. Bulgarelli, F. Fuschino, F. Gianotti, M.
Trifoglio (INAF/IASF, Bologna), F. Longo, C. Pontoni, M. Prest, E.
Rotta, E. Vallazza (INFN - Sez. Trieste), P. Bresciani, N.
Kocjancic, D. Rossi, R. Starec, R. Terpin (MIPOT, Cormons), Z.
Durna, J. Vincenz (ILFA Gmbh, Frankfurt), S. Mikkelsen, E.
Nygaard, A. Suleyman, B. Sundal, K. Yoshioka (IDE AS, Oslo),
 L. Acquaroli, M. Angelucci, M. Angelucci, F. Antichi,  G. Babini,
 M. Calabrese, B. Morelli, G.
Pedrazzoli, U. Ricciardi, S. Sideri (Oerlikon Contraves Italia),
E. Artina, P. Bastia, A. Bonati, G. Cafagna, L. Maltecca, G.
Marseguerra, F. Monzani, L. Nicolini, M. Patane', R. Pavesi, P.
Radaelli, F. Ticozzi, V. Vettorello, E. Zardetto (Thales-Alenia
Space Italia, formerly Laben, Milan), S. Alia, G. Annoni, E.
Collavo, I. Ferrario, B. Garavelli, M. Giacomazzo, S. Legramandi,
A. Longoni, C. Maini, P. Sabatini (Carlo Gavazzi Space, Milan),
L.A. Antonelli, G. Fanari, P. Giommi, C. Pittori, B. Preger, R.
Primavera, P. Santolamazza, S. Stellato, F. Verrecchia (ASI
Science Data Center).

\textbf{FIGURE CAPTIONS}

FIGURE 1: A sketch of the exploded SA structure.\\

FIGURE 2: Photographs of one of the four flight units of the SAFEE
   boards, taken during the integration process in 2005. The photo on
   the left shows both the horizontal and vertical boards (here open),
   connected by the flexible section of the PCB. The photo on the right
   shows the integration process with the detector, with the vertical
   section closed in its aluminum box.\\

FIGURE 3: The flight unit of the SA Detection Plane after its
   mechanical integration in MIPOT.\\

FIGURE 4: The flight unit of the carbon-fiber/tungsten SA
collimator.
   The bottom view allows
   to see the separation between collimator cells and the copper plates
   providing the electrical grounding of the structure.\\

FIGURE 5: A simulated SA exposure to a blank sky (i.e., no point
   sources), with the diffuse X-ray
   background partially blocked by the Earth. The top panel shows the
   detector images of the two homologous detectors. The mid panel shows
   the corresponding sky images, with the significant effect of the
  background inhomogeneity. The bottom panel is the sum of the two
  sky images, showing their compensation, due to the mask/anti-mask
  configuration of the two homologous detectors.\\

FIGURE 6: The flight unit of the SA mask-collimator system.
   The four individual coded masks
   are visible, as well as the sections at the outer edges where the sample
   strips were manufactured and stripped off for the quality control.
   The 3 optical cubes used for alignment purposes are also visible on
   the corners.\\

FIGURE 7: SA in its flight configuration, integrated with the
AGILE
   payload during the integration campaign in at the CGS facilities
   in Tortona.\\

FIGURE 8: Gain (ADC channel per fC) for all the 6144
   individual channels of SuperAGILE, as measured on March 2007.\\

FIGURE 9: Energy resolution (keV) for all the 6144
   individual channels of SuperAGILE, as measured on March 2007.\\

FIGURE 10: Analog energy threshold setting (keV) for all the 6144
   individual channels of SuperAGILE, as measured on December
   2006, with the launch configuration setting.\\

FIGURE 11: Energy spectrum of the background in the D3 unit, with
   a nominal threshold setting at 17 keV. The energy units are given
   in Pulse Height Amplitude units, each one approximately corresponding
   to 1 keV.\\

FIGURE 12: Energy spectrum of a Cd$^{109}$ radioactive source
obtained with
   the SA unit D0, placing the source on axis at a distance of $\sim$200 cm.
   The energy units are given
   in Pulse Height Amplitude units, each one approximately corresponding
   to 1 keV.\\

FIGURE 13: Energy spectrum of a I$^{125}$ radioactive source
obtained with
   the SA unit D0, placing the source on axis at a distance of $\sim$200 cm.
   The energy units are given
   in Pulse Height Amplitude units, each one approximately corresponding
   to 1 keV.\\

FIGURE 14: Zoom-in of the images of the collimated beam from the
X-ray
  tube, for an incidence angle near to the instrument optical axis
  and energy selected in the range 27-45 keV. Top left is unit D0, top right
  D1, bottom left D2 and bottom right D3. The angular size of each bin
  in these images is 3 arcmin.\\

FIGURE 15: Zoom-in of the images of the collimated beam obtained
with unit
  D0 from the X-ray
  tube, for an incidence angle near to on-axis (left), $\sim$10$^{\circ}$
  (center) and $\sim$20$^{\circ}$ (right) off axis
  and energy selected in the range 27-45 keV.  The angular size of each bin
  in these images is 3 arcmin.\\

FIGURE 16: Same images as in Figure \ref{tube_offaxis_zoom}, but
  showing the full field of view, allowing to outline the image
  distortions due to the incomplete sampling of the mask code at
  off-axis locations. As above, incidence angle is near to on-axis (left),
  $\sim$10$^{\circ}$ (center) and $\sim$20$^{\circ}$ (right) off axis
  and energy is selected in the range 27-45 keV.  The angular size of each bin
  in these images is 3 arcmin.\\

FIGURE 17: Reconstructed images of a Cd$^{109}$ radioactive source
(22 keV)
  at 205 cm
  distance from SA, on axis (left) and 30$^{\circ}$ off-axis (center).
  On the right, the image of a Cs$^{137}$ source
  (32 keV) on axis, at the same distance.
  The angular size of each bin in these images is 3 arcmin.\\

FIGURE 18: Reconstructed images of a Cd$^{109}$ radioactive source
(22 keV)
  at $\sim$200 cm distance from SA, for off-axis angles from -40$^{\circ}$
  to +50$^{\circ}$, step 10$^{\circ}$ (from top left to bottom right).
  The angular size of each bin in these images is 3 arcmin. \\

FIGURE 19: Effective area of one of the 4 SA Detection Units,
derived by
  means of analytical calculations, as
  a function of energy, at different off-axis angles.\\

FIGURE 20: \textit{Left:} Monte Carlo simulation of the 50 ks
sensitivity of 2 SA
  Detection Units, as a function of the source position in the field
  of view, in the 15-45 keV energy range. The sensitivity is expressed
  in terms of mCrab units. \textit{Right:} combined sensitivity (bolean AND)
  of the 4 SA detectors. The given sensitivity in each point
  is the one relevant to the
  least sensitivity of the two pairs of detectors.\\

FIGURE 21: Estimation of the statistical SA Point Source Location
Accuracy
  (here D0 unit) in terms of the distribution of the positions of the
  barycenter in the reconstructed image, in a sample of statistically
  independent measurements performed
  with a 22 keV source at 200 cm distance, during the ground
  calibration. Panels on the left and rights show the measurement with a signal-to-noise
  ratio of $\sim$5 and $\sim$10, respectively.\\

FIGURE 22: \textit{Top}: Sky image of the on-axis point-like
source in ideal conditions with absolutely stable pointing.
\textit{Second}: Sky image of the same source without attitude
correction. Source shape roughly matched the pointing error
histogram for a given direction. In the case shown the pointing
errors had a trend towards positive coordinates, thus making the
uncorrected image to be centered in the wrong position.
\textit{Third}: The same as above, but after applying the
on-ground attitude correction method. Note the decreasing of the
coding noise in the image for angles above 30 degrees compared
with the ideal case. \textit{Bottom}: The same as above, but after
applying the simplified on-board attitude correction method.
Correction parameter was set for the best correction at 20 degrees
off-axis in the field of view.\\

FIGURE 23: \textit{Left}: Simulated SA observation of the Galactic
Center
   region, as described by second INTEGRAL Catalog (Bird et al. 2006).
\textit{Right}: Sources identified in the same image by the SA
IROS procedure after 10 iterations. \\

\end{document}